\documentclass[journal]{IEEEtran}

\usepackage{ifpdf}
\usepackage[latin1]{inputenc}
\usepackage{graphicx}
\usepackage{epsfig}
\usepackage{amssymb,amsmath,array}
\usepackage{epstopdf}

\ifCLASSINFOpdf
\else
\fi

\def\um{\ensuremath{\mu {\mathrm{m}}}}

\def\TeV{\ifmmode {\mathrm{\ Te\kern -0.1em V}}\else
                   \textrm{Te\kern -0.1em V}\fi}%
\def\GeV{\ifmmode {\mathrm{\ Ge\kern -0.1em V}}\else
                   \textrm{Ge\kern -0.1em V}\fi}%
\def\MeV{\ifmmode {\mathrm{\ Me\kern -0.1em V}}\else
                   \textrm{Me\kern -0.1em V}\fi}%
\def\keV{\ifmmode {\mathrm{\ ke\kern -0.1em V}}\else
                   \textrm{ke\kern -0.1em V}\fi}%
\def\eV{\ifmmode  {\mathrm{\ e\kern -0.1em V}}\else
                   \textrm{e\kern -0.1em V}\fi}%

\let\gev=\GeV

\begin{document}
\pdfoutput=1
%
\title{Physical limitations to the spatial resolution of solid-state detectors}
\author{M.~Boronat, C.~Marinas, A.~Frey, I.~Garcia, B.~Schwenker, M.~Vos, F.~Wilk
\thanks{This work was supported by the German Federal Ministry for Research BMBF, the VolkswagenStiftung and the Spanish Ministerio de Ciencia e Inovaci\'on. The studies reported in this paper are based on data collected on prototypes constructed by the DEPFET collaboration. The authors furthermore acknowledge the support and infrastructure of the AIDA project (EU FP7 grant agreement 262025) for beam tests at CERN and DESY. We thank prof. Hans Bichsel for help accessing and understanding the predictions of his model, Hans-G\"unther Moser and Rainer Richter for very useful comments to the manuscript and Erik Heijne for triggering our interest in the limitations of solid state devices and for his helpful suggestions to the manuscript.
 
C. Marinas is with the University of Bonn, D-53115 Bonn, Germany.
 
M. Boronat, I. Garc\'ia and M. Vos are with IFIC (UVEG/CSIC), E-46980 Valencia, Spain (e-mail: marcel.vos@ific.uv.es).
 
A. Frey, B. Schwenker and F. Wilk are with G\"ottingen University, D-37077 G\"ottingen, Germany.

}
}
\markboth{Draft to be submitted to Transactions on Nuclear Science, 2013}%
{Shell \MakeLowercase{\textit{et al.}}: }
\maketitle

\begin{abstract}
In this paper we explore the effect of $\delta$-ray emission, fluctuations in the signal deposition on the detection of charged particles in silicon-based detectors. We show that these two effects ultimately limit the resolution that can be achieved by interpolation of the signal in finely segmented position-sensitive solid-state devices.
\end{abstract}


\IEEEpeerreviewmaketitle

\section{Introduction}

Since the advent of silicon vertex detectors in the 1980s an 
intense detector \mbox{R \& D} program has led to a continuous improvement
of many aspects of the performance of silicon-based devices for charged
particle detection. Today's detector are much more radiation-hard than 
their precursors. The development of pixel detectors has increased the
detector granularity that can be attained by several orders of magnitude,
allowing for robust pattern recognition in a dense environment.
The installation of approximately 100 $m^2$ of silicon $\mu$-strip and
 at the LHC experiments mark a milestone in the history of solid-state 
devices for charged-particle detection.

The key to the position-sensitivity of silicon-based detectors is
the segmentation of the silicon wafer - the pixel size or the micro-strip
pitch. The spatial resolution of the device is related to dimension of 
the read-out segments, be it the pixel size or the micro-strip detector pitch.  
If the signal is collected in a single strip or pixel the 
residuals (measured position minus true position) form a uniform 
distribution with an extension equal to the pitch $p$~\cite{Campabadal:2005cn}. 
The resolution $\sigma$, measured as the root-mean-square of the residual 
distribution, then has the following simple relation to the pitch: 
$\sigma = p/\sqrt{12}$. A spatial resolution that is significantly 
better than this 'binary limit' can be obtained by interpolating the 
position on the basis of the signal measured on two neighbouring strips 
or pixels. For detectors where the signal-to-noise ratio (S/N) 
of the signal amplitude measurement on each cell dominates the resolution
(the exact range of validity is discussed in detail in the following) 
the resolution is given by a simple formula~\cite{Turchetta:1993vu}: 
\begin{equation}
\sigma \propto \frac{p}{S/N}
\label{eq:resolution}
\end{equation}
In practice, the signal $S$ is often
measured as the Most Probable Value of the distribution of the signal
of Minimum Ionizing Particles (MIPs) of clusters of pixels or strips
and $N$ is identified with the noise of a single cell. 

This 'charge sharing' underlies the performance of the most precise
$\mu$-strip detectors. The review in Reference~\cite{Beringer:1900zz} 
quotes a typical 
resolution for $\mu$-strip detectors equal to the read-out pitch divided 
by a factor 3 to 7. With a sufficiently large $S/N$ ratio the gain in 
spatial resolution can be larger than that. Several groups have produced 
devices with a spatial resolution of a few \um{} 
already a long time ago~\cite{Antinori:1990hz, Straver:1994it}. 
This precision is by no means limited to a single coordinate. 
Pixel detectors achieve a similar resolution (below 5~\um{}) simultaneously 
on two coordinates~\cite{Abe:1997bu,Akiba:2011vn, Winter:2012ms}. 
Recently, several groups have demonstrated a resolution that approaches 
a single~\um{}~\cite{Andricek:2011zza, Velthuis:2008zza, Battaglia:2011vi}.

We expect the miniaturization of structures in silicon-based devices 
to continue its current rapid progress. At the same time, 
techniques for integrating the Front End electronics with the 
active sensing element are bound to become more sophisticated. 
It is then natural to
expect that in the near future silicon detectors can be produced with 
very fine segmentation ($p <<$ 10 \um) and a much improved $S/N$ ratio
in comparison to today's state-of-the-art position sensitive devices. 
It is likely, therefore, that technology should be available in the
not-so-distant future that, according to equation~\ref{eq:resolution}, 
should yield resolutions well below 1~\um{}.

For today's resolutions of 5-10 \um{} the assumption underlying the operation
of silicon detectors - that the position
of the center-of-gravity of the cloud of charge carriers released by
the ionizing particle can be identified with the position of the
impinging particle - is certainly true to good approximation. 
But, is that still the case if we require a resolution of 100 nanometers?
Or, to phrase the question more generally:
\begin{quotation}
{\it What is the ultimate position resolution that can be obtained with 
solid-state devices that rely on charge sharing between neighbouring cells?}
\label{question1} 
\end{quotation}

In this paper we explore limitations that stem from the physical 
process responsible for the generation of a signal when charged particles
traverse a thin layer of silicon.
We use measurements in beams of particles at CERN 
and DESY to quantify these limitations. The response of DEPFET active pixel 
detectors~\cite{Velthuis:2008zza,Kemmer:1986vh,Alonso:2012ss} 
with excellent S/N ratio (well over 100) and small pixel size 
(down to 20 $\times$ 20~\um$^2$) is characterized using a very precise
beam telescope based on thin MIMOSA sensors made available within the
AIDA project~\cite{aida}. For a detailed description of the setup and analysis
the reader is referred to Ref.~\cite{Andricek:2011zza}
The resolution measurements based on these data are compared
to GEANT4~\cite{Agostinelli:2002hh,Allison:2006ve} simulations
with a detailed description of the DEPFET response~\cite{Drasal:2011zz}.
We note that even if the measurements correspond to a particular
detector technology, the conclusions apply quite 
generally to position-sensitive devices based on silicon. 

In Section~\ref{sec:landau} the energy deposition by minimum ionizing particles
in thin layers of silicon is briefly reviewed and the straggling functions
used in the following Sections are presented. 
Next, in Section~\ref{sec:deltas} we briefly summarize the findings
of a detailed study into the rate and range of $\delta$-rays - electrons 
knocked out of their shell by the impinging particle. 
The next two Sections explore the limitations to the spatial
resolution. Section~\ref{sec:perp} presents the results of a study 
of the resolution that can be achieved for particles under perpendicular 
incidence. In Section~\ref{sec:angle} we consider position measurements 
for particles traversing the silicon detector 
under an angle.
Finally, in Section~\ref{sec:conclusion} we summarize the findings
and present an outlook.

\section{Straggling functions}
\label{sec:landau}

The energy deposited by charged particles in thin layers of silicon is 
described by straggling functions. In high energy physics 
these are commonly referred to as {\em Landau} distributions, even if 
the distributions differ substantially from the predictions of Landau's 
original model. Here we use the model by H. Bichsel~\cite{bichsel}.

\begin{figure}[h!]
 \centering 
  \includegraphics[width=0.9\columnwidth]{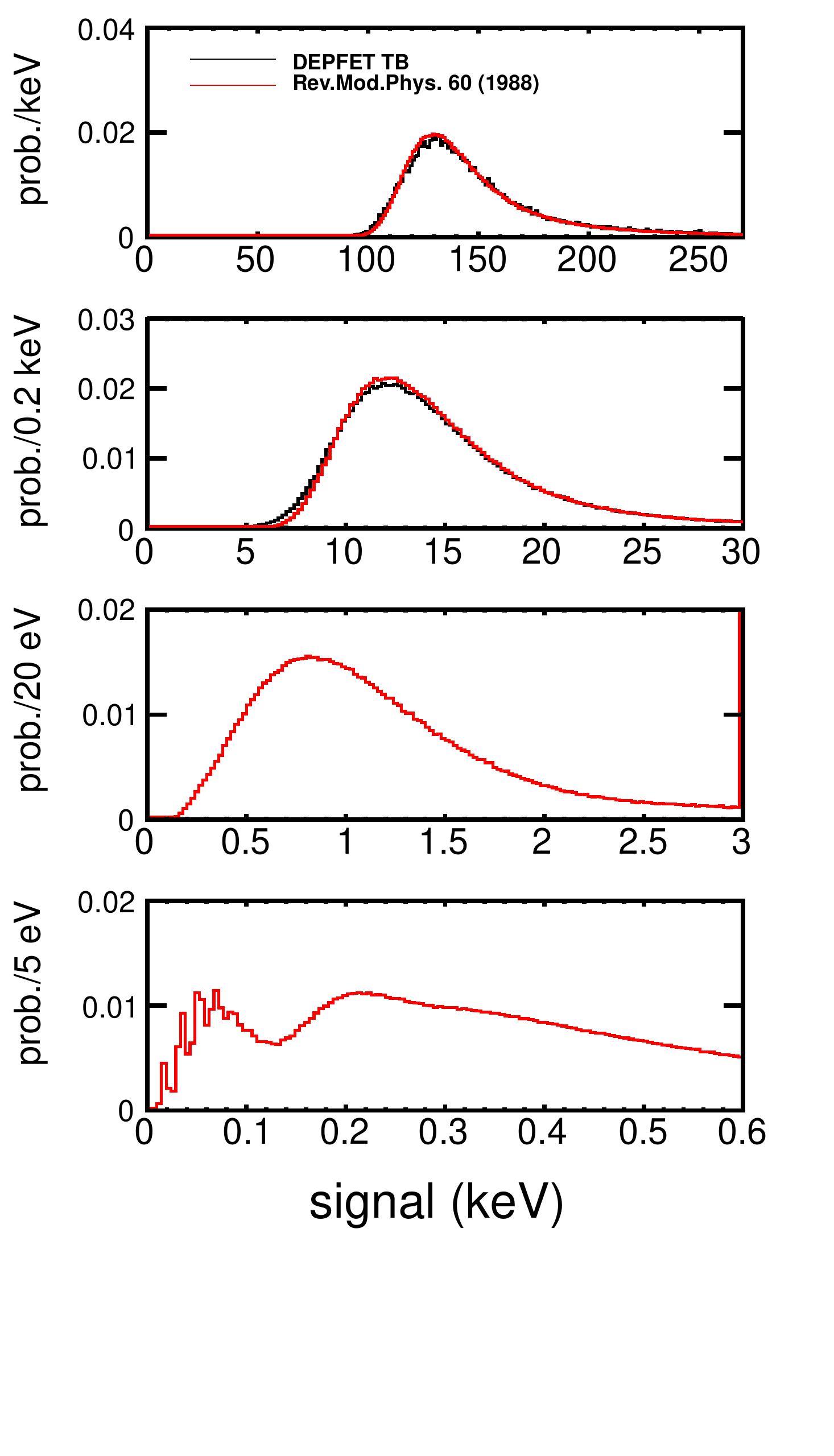}
\vskip-13ex
 \caption{Distribution of the energy deposition in keV by minimum ionizing particles traversing silicon sensor sensors with thicknesses of 450 \um{} (top panel), 50 \um{} (second panel from the top), 5 \um{} (third panel from the top) and 1 \um{} (bottom panel). The dashed (red) lines represent the prediciton of the model of H. Bichsel~\cite{bichsel}. Where available, the measured distribution has been overlaid, with a calibration that makes sure that the Most Probable Value of data and prediction agree.}
\label{fig:landaus}
\end{figure}

The predictions for a few sample thicknesses ranging from nearly
half a millimeter to 1 \um{} are presented in Figures~\ref{fig:landaus}.
Experimental distributions obtained with DEPFET sensors~\cite{Andricek:2011zza} are superimposed for the curves corresponding to a sensor thickness of 
450~\um{} and 50~\um{}. The excellent agreement of the data confirms the 
accuracy of these predictions for sensor thicknesses $d$ ranging from 
several tens to several hundred~\um. The distributions for $d =$ 50~\um{}
and $d=$ 450~\um{} are moreover found to be in good agreement~\cite{thesisbenjamin} with the predictions of the simplified energy loss model~\cite{LassilaPerini:1995np} in GEANT4~\cite{Agostinelli:2002hh,Allison:2006ve}, that assumes that atoms have only two energy levels and ionization energy loss is distributed according to a one over energy squared law.

The results in Figure~\ref{fig:landaus} clearly show that fluctuations 
in the energy deposition become more prominent as the thickness of 
the sensors decreases. The width (Full Width at Half Maximum, FWHM) 
divided by the position of the peak yields 0.32 and 0.62 for 
thicknesses of 450~\um{} and 50~\um{}, respectively. The ratio 
approaches 1 for very thin layers. We discuss the implications
of these fluctuations in Section~\ref{sec:angle}.

\section{$\delta$-electrons}
\label{sec:deltas}

Charged particles traversing material ionize atoms along the trajectory. 
Most free charge carriers are formed at a very small distance from 
the particle trajectory and in typical detectors signal sharing between 
neighbouring cells is dominated by diffusion of the charge 
carriers as they drift to the read-out plane. 

\begin{figure}[h!]
 \centering 
  \includegraphics[width=0.7\columnwidth]{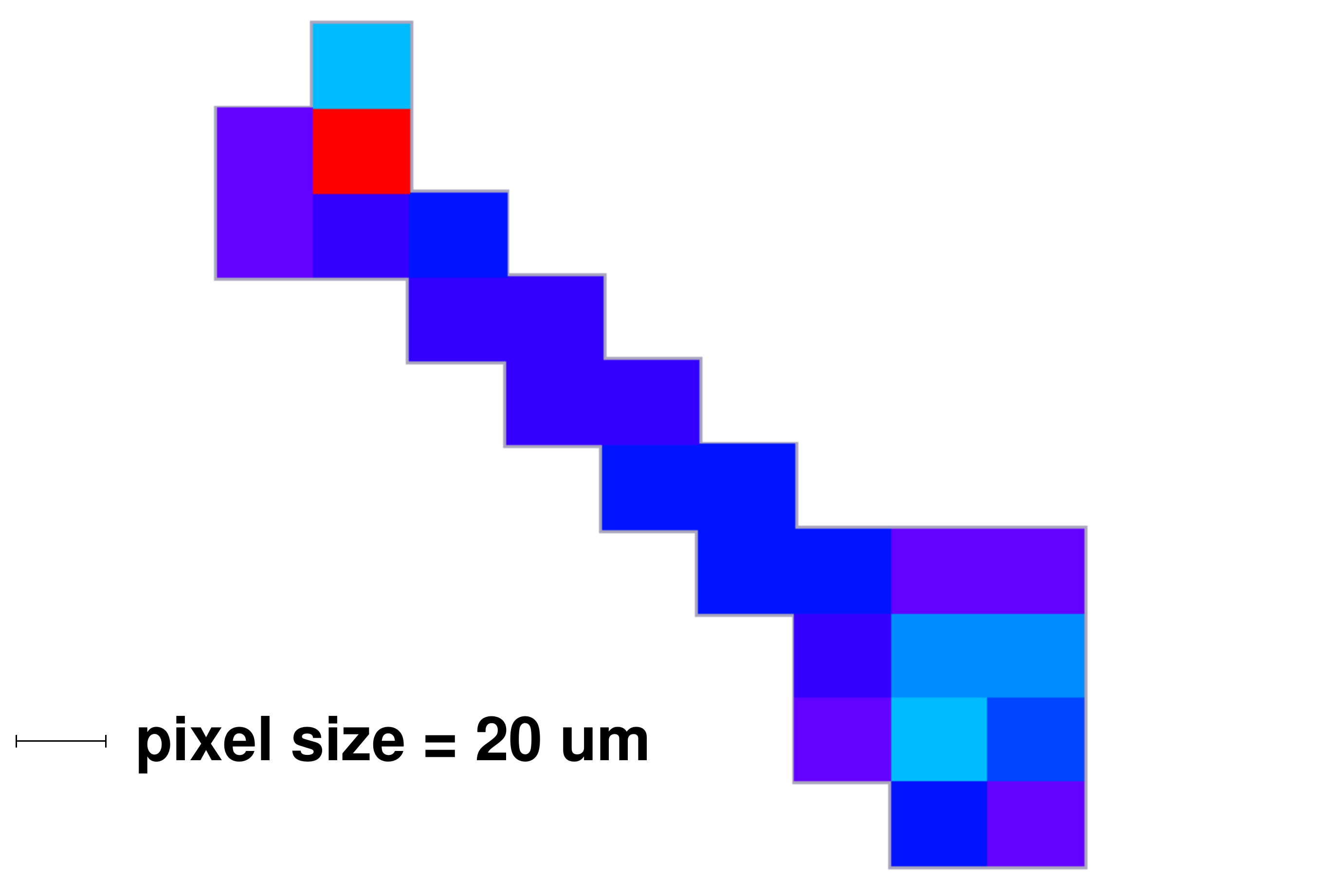}
 \caption{A $\delta$-electron candidate registered in a DEPFET pixel detector 
with 20~\um{} pitch and 450~\um{} thickness registered in a beam test 
with 120~\gev{} pions at the CERN SPS. Pixels with a signal above the noise
threshold are indicated as squares. The colour coding indicates the signal
height (blue for the smallest signal, red for the largest signal). }
\label{fig:delta}
\end{figure}

Occasionally, however, the momentum transfer between the incident 
particle and an electron in a silicon atom is large enough that 
a secondary track is formed. 
An image of a $\delta$-electron candidate registered in a DEPFET pixel
detector is shown in Figure~\ref{fig:delta}. The primary cluster, given by the 
position of incidence of the pion predicted by the reference telescope, 
is located in the upper, leftmost corner. The long tail diagonally
across the display and the secondary clusters in the lower, rightmost 
corner are interpreted as a $\delta$-electron that travels 
a considerable distance (more than 100~\um{} in this case) 
before it stops releasing its remaining energy.
Obviously, the center of gravity of the signal in this event is a poor 
estimate of the position of the primary pion.

\begin{figure}[h!]
 \centering 
  \includegraphics[width=\columnwidth]{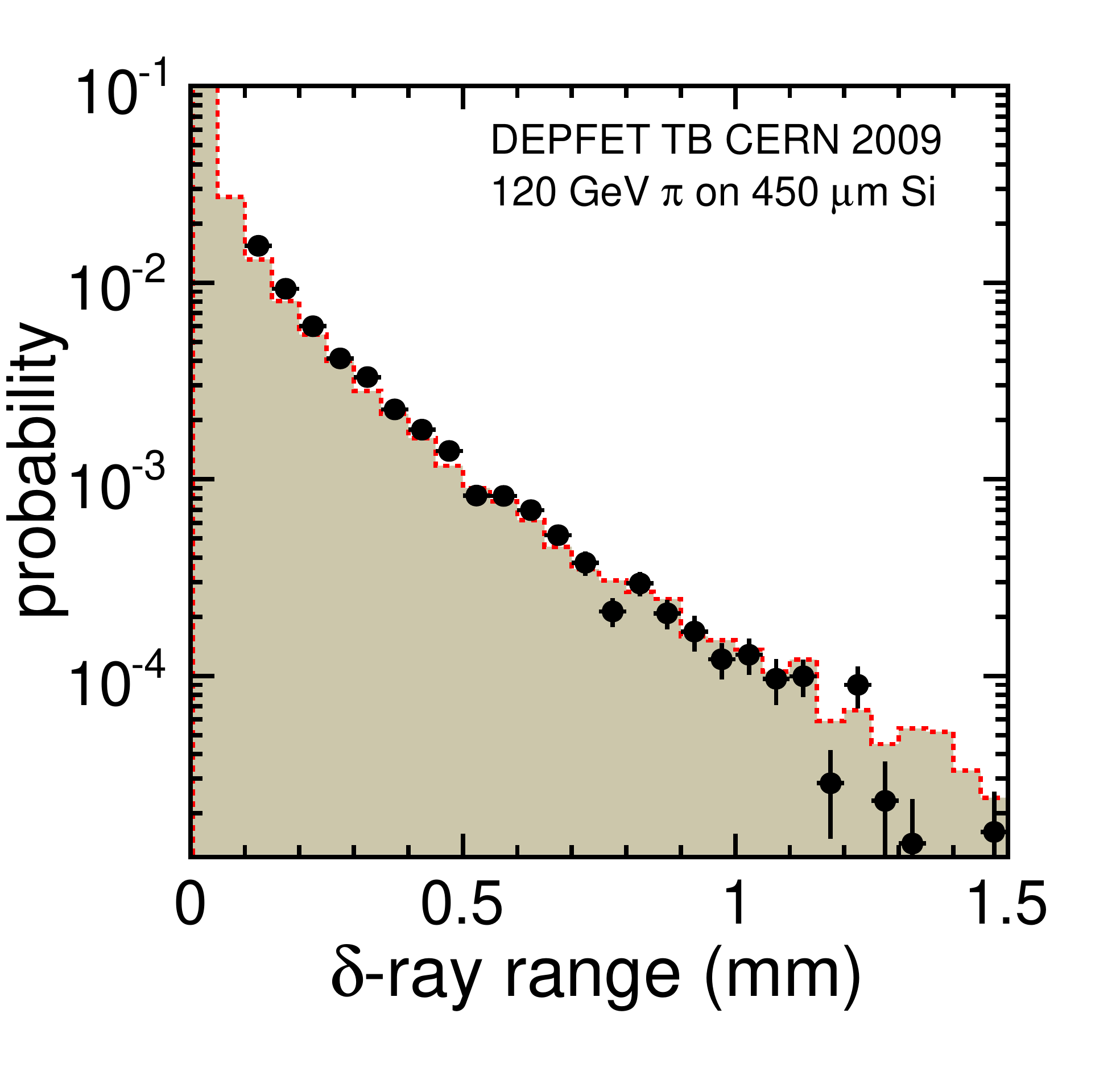}
 \caption{The probability for long-range $\delta$-electrons (range 
perpendicular to the particle trajectory $>$ 100~\um{}) to be emitted 
when a 120~\gev{} pion traverses a 450~\um{} thick DEPFET detector. 
The data (solid black points) are compared to a GEANT4 prediction (dashed 
histogram).}
\label{fig:deltarange}
\end{figure}

The $\delta$-electron emission rate has been measured from beam test data.
Figure~\ref{fig:deltarange} presents the probability that a 120~\gev{} 
pion emits a $\delta$-electron when traversing a 450~\um{} thick silicon 
sensor versus the range in the direction perpendicular to the particle 
trajectory. To isolate long-range secondaries, the threshold of the DEPFET 
devices is set to three times the pixel noise (the average single noise
is equivalent to less than 300 electrons) and large clusters, with at 
least 7 pixels above threshold, are selected. 
Nuclear interactions are vetoed by requiring a high-quality track
to leave a signal in several layers - including one downstream of the device
under consideration - of the experimental setup. The result is corrected
for the detection efficiency of $\delta$-electrons, estimated with GEANT4.

The curve that overlays the data points corresponds to the prediction 
of GEANT4. For energy depositions above a certain energy threshold GEANT4
simulations $\delta$-electron explicitly as a secondary particle that is
tracked through the GEANT4 volumes and suffers energy loss and multiple Coulomb 
scattering. Softer particles are simply accounted 
for in the continuous ionization energy loss. We used a range cut $r =$ 1~\um{} in silicon. With this choice the production rate for long-range 
$\delta$-electrons is found to be in good agreement
(see also Reference~\cite{ATLAS:2013vma}).  
Further details are found in Ref.~\cite{goettingen}. 
 
Fortunately, spectacular events such as that of Figure~\ref{fig:delta} are 
relatively rare. After correction for the $\delta$-ray reconstruction 
efficiency we find a 5.4\% probability for a secondary track with 
a range in the plane perpendicular to the particle trajectory of at 
least 100~\um{}, in qualitative agreement with the findings of 
Ref.~\cite{heijne,Field}. Shorter-range $\delta$-electrons are, however, 
exceedingly common. So common, in fact, that they have a considerable 
impact on the spatial resolution of the most precise devices. 
In the next Section we evaluate this impact.

\section{Perpendicular incidence}
\label{sec:perp}

The impact of $\delta$-electrons on the spatial resolution 
is evaluated using data taken with a DEPFET device with a S/N 
ratio in excess 100 in a beam of 120~\gev{} pions. 
The resolution analysis described in Ref.~\cite{Andricek:2011zza} compares 
the response of several read-out planes forming a precise beam telescope.
A single-point spatial resolution of 1.4~\um{} is extracted.

The evolution of the spatial resolution with S/N ratio is evaluated adding
random noise to the response of the DEPFET pixels. 
We create {\em pseudo-measurements} for a device with poorer S/N 
by smearing the pixel response using a random number 
with a Gaussian distribution centered at 0 and with a variable 
width. The spatial resolution analysis 
is repeated for 64 values of this 'noise, corresponding to 
S/N ratios between 10 and 115.

\begin{figure}[h!]
 \centering 
  \includegraphics[width=\columnwidth]{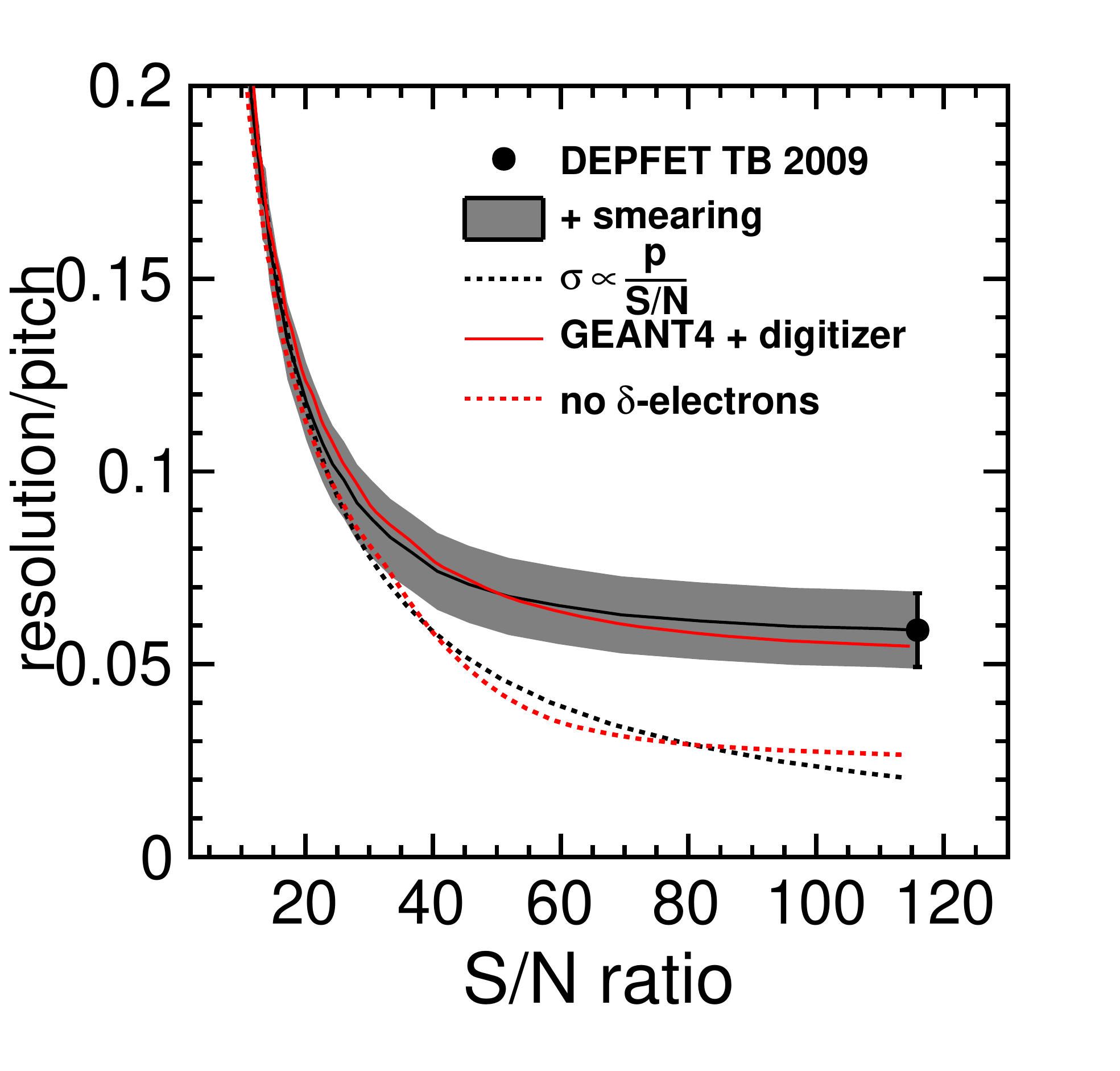}
 \caption{The spatial resolution divided by pixel size for 120~\gev{} pions 
from the CERN SPS perpendicularly incident on a 450~\um{} thick DEPFET device
with a 24~\um{} pixel size. The data point indicated by a solid marker 
corresponds to the nominal measurement. The red curve with the grey error
band is obtained by smearing the signal. The result is compared to the
expected evolution of the resolution according to 
Equation~\ref{eq:resolution} and to a GEANT4 simulation 
of the setup, with and without $\delta$-electron emission.}
\label{fig:delta_plot}
\end{figure}

The result is presented in Figure~\ref{fig:delta_plot}. The measurement
on the test beam data is indicated with a marker. The resolution obtained
with smeared data are represented by the curve with error band. 
For S/N ratios between 10 and 20 the evolution of the resolution 
follows the behaviour predicted by Equation~\ref{eq:resolution}, 
indicated by the dashed curve in the figure (the proportionality constant
is fixed to 2.3, such that the integral of the curve over the S/N interval
from 20 to 50 agrees with that of the simulated curve without 
$\delta$-electron emission, that is explained below).
For S/N ratio larger than 20 the resolution determined 
on data is found to fall behind the predicted evolution. Between 
values of the S/N ratio of 60 and 120 the spatial resolution
should improve by a factor 2 according to Equation~\ref{eq:resolution}. 
Instead, we observe a mere 15\%. Obviously, the detector resolution does
not continue to scale with the inverse of the S/N ratio in this regime.

The result of a GEANT4 simulation of the setup with a detailed {\em digitizer}
model~\cite{Drasal:2011zz,b2tdr,thesisbenjamin} of the DEPFET detector 
response is superposed on the
data curve in Figure~\ref{fig:delta_plot}. The simulation includes 
the effect of $\delta$-electrons, but does not take into account 
imperfections in the read-out (such as non-uniformities in the gain)
or measurement errors (such as an incorrect treatment of the finite telescope
resolution or residual misalignment of the setup). An 
adequate description is obtained of the observed evolution of the
resolution with increasing S/N ratio. The difference in resolution
between data and Monte Carlo is 100 nm in case no smearing is applied, 
and less everywhere else. The dashed red line indicates the 
resolution vs. S/N ratio curve for the same simulation 
without $\delta$-electron emission. In this study the position is
obtained by interpolation of the observed signal assuming linear
signal sharing between pixels (the center-of-gravity method). This is found to
be an adequate approximation for data and Monte Carlo samples; application
of the $\eta$ correction to take into account non-linearity does not lead
to a significant improvement of the result. This does not hold, however, 
for the simulation without $\delta$-electron emission, where 
the $\eta$ correction has an important effect. The red dashed curve therefore
corresponds to the resolution obtained after the $\eta$ correction. 
The slope at large S/N ratio
is restored, yielding resolutions at large S/N ratio that are clearly 
incompatible with the data and come close to the evolution predicted by
Equation~\ref{eq:resolution}. These findings confirm that 
$\delta$-electrons play an important role, limiting the resolution 
for detectors with very large S/N ratio.

\section{Incidence under an angle: signal fluctuations}
\label{sec:angle}

The discussion of $\delta$-electrons above applies to particles under
an arbitrary angle. For non-perpendicular incidence an additional
effect must be considered. In this case fluctuations in the energy 
deposited by charged particles along their trajectory through 
the sensor are known to affect the position. The same is true 
for operation in a magnetic field, as charge carriers drifting 
to the read-out plane of the sensor are deviated by the magnetic 
field, rendering the situation equivalent to incidence under 
an angle, the Lorentz angle, given by $\tan{\theta_L} = \mu_H B$, where
$\mu_H $ is the Hall mobility of the charge carriers and $B$ the
magnetic field in Tesla. In this Section we quantify the impact of 
these 'Landau' fluctuations. 
We consider the situation where the projection on the read-out 
plane of the particle trajectory through 
the sensor has a length equal to the pitch $p$.


For particles crossing the sensor at an angle generally the fraction
of the signal deposited in each cell is proportional to the path length through 
the silicon in that cell {\em on average}. On an event-by-event basis, 
however, the amount of signal deposited in each segment fluctuates strongly.
The signal sharing between the neighbouring cells therefore does not allow
to estimate the position of the incoming particle to arbitrary precision, no
matter how good the S/N ratio of the device.

\begin{figure}[h!]
 \centering 
  \includegraphics[width=\columnwidth]{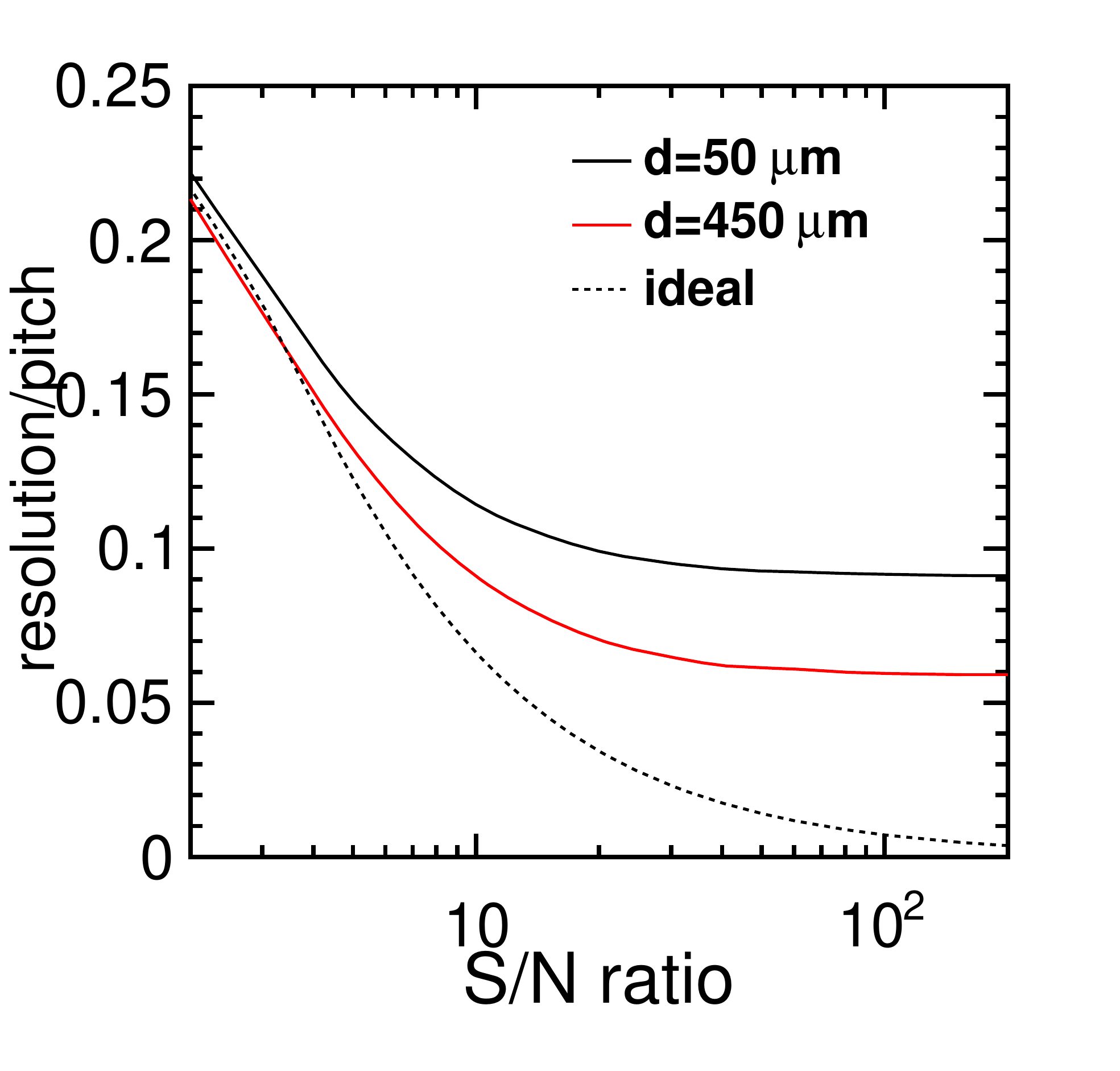}
 \caption{The spatial resolution divided by pitch versus S/N ratio. 
The dashed curve corresponds to an ideal detector where the 
signal is proportional to path length. The two solid curves 
include the effect of fluctuations following the straggling functions.}
\label{fig:fluctuations_simple}
\end{figure}

In Figure~\ref{fig:fluctuations_simple} we estimate the impact 
of {\em Landau} fluctuations for different sensor thicknesses using 
a toy Monte Carlo calculation. 
The dashed line corresponds to an ideal detector where the signal in 
each cell is strictly proportional to the path length $l$ of the 
particle through that particular cell  ($S \propto l$).
The other two curves correspond to more realistic setups corresponding
to detector thickness of 50~\um{} and 450~\um{}. On average the signal
is still proportional to the path length, but now fluctuations are
simulated according to the straggling functions in Figure~\ref{fig:landaus}. 
As expected the resolution of the ideal detector improves with the inverse of
the S/N ratio. For a S/N ratio of 100 the ideal detector would 
reach a resolution of 1\% of the pitch.

The resolution of the realistic detectors is very similar to that of the
ideal detector for poor S/N ratio, where the fluctuations in the signal
are negligible in comparison to the read-out noise. For a S/N ratio 
of 10 the differences between ideal detector and the more realistic simulation
are already quite significant; the ideal detector achieves a resolution
of 6\% of the pitch (five times better than the binary limit), whereas
the realistic detectors have resolutions ranging from 9\% of the pitch for
very thick devices to 12\% for a thickness of 50 \um{}.
For large S/N ratio, of approximately 30 for
the thicknesses considered here, the resolution of the realistic model
reaches an asymptotic value. 
At that point the {\em Landau} fluctuations form the dominant
signal distortion, more important than the read-out noise. A further increase 
of the S/N ratio does not lead to an improved spatial resolution. 
For thick devices fluctuations are relatively small and an asymptotic
resolution of 5\% is attained. The thinnest device considered here,
with $d=$ 50~\um{} cannot attain a resolution beyond 9\%.

\begin{figure}[h!]
 \centering 
  \includegraphics[width=\columnwidth]{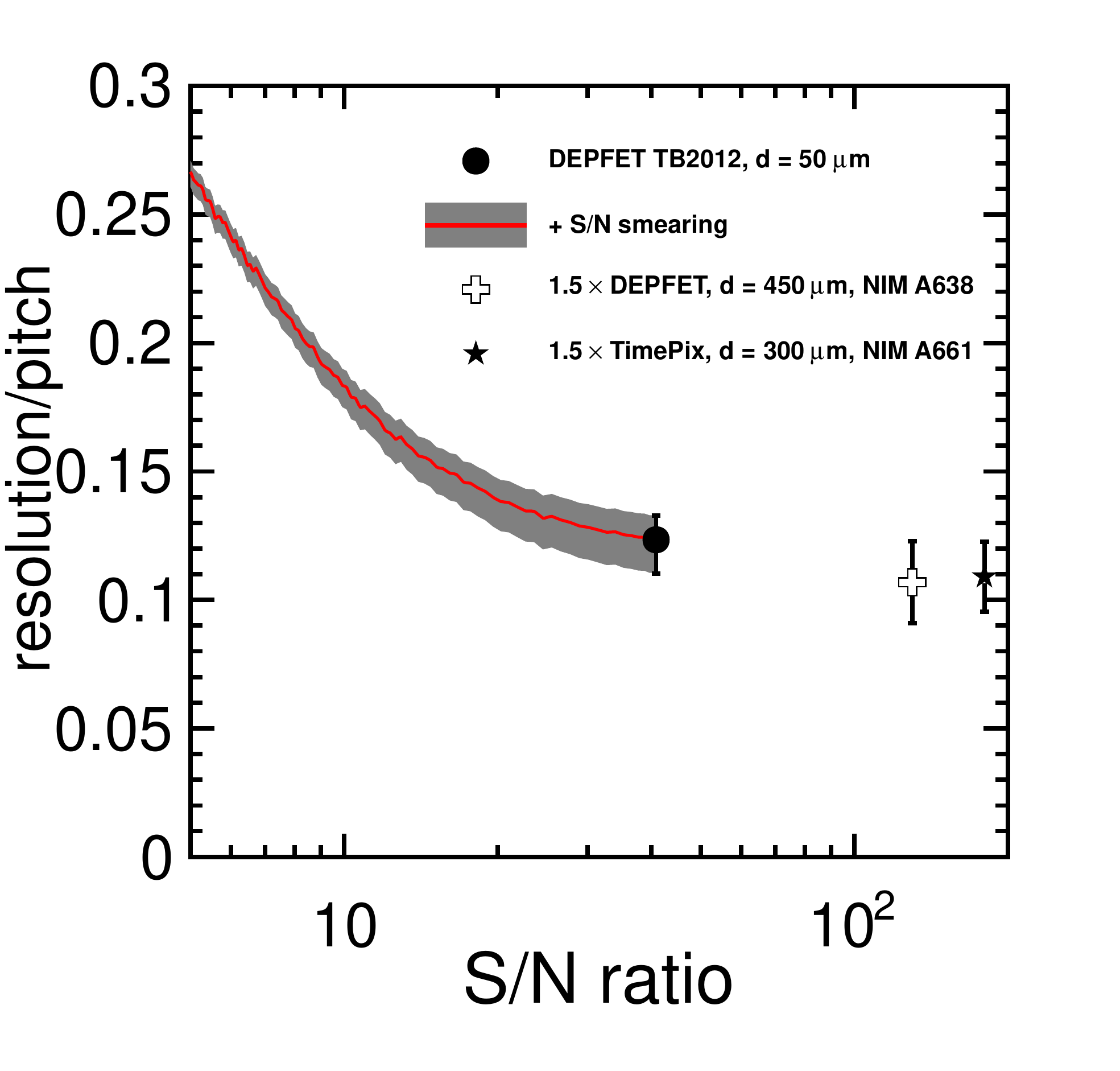}
 \caption{The spatial resolution divided by pitch for 3.75~\gev{} electrons
traversing a 50~\um{} thick DEPFET device with 50~\um{} pitch 
under a 45$^\circ$ angle versus the S/N ratio (filled circular marker). 
The resolution results for a S/N ratio below 40 are obtained 
by smearing the signal (red curve with grey error band). }
\label{fig:fluctuations_data}
\end{figure}

An empirical demonstration of the impact of signal fluctuations on
the spatial resolution is obtained in a beam of particles under a 45$^\circ$ degree angle. The measurement setup is identical to that described in 
Reference~\cite{Andricek:2011zza}. The prototype under test is a 50~\um{} 
thick DEPFET device with a pixel 
size of 50~\um{}. Further information and details on the methodology to extract
the spatial resolution are found in Reference~\cite{thesisbenjamin}.
The spatial resolution divided by the read-out pitch is plotted versus S/N 
ratio in Figure~\ref{fig:fluctuations_data}. Values of the S/N ratio below 
40 are obtained by smearing the signal using the method introduced in 
Section~\ref{sec:perp}.
Two further measurements on devices with large S/N ratio from 
references~\cite{Akiba:2011vn,Andricek:2011zza} are plotted for
comparison. In each
case the incidence angle is such that the length of the projection of
the particle trajectory on the read-out plane is equal to the pixel size.
To compare these three devices with different thicknesses on an equal 
scale the result has been multiplied by a factor 1.5, obtained from the
simulation discussed previously. 
The resolution expressed in terms of the pixel size is $\sigma/p \sim$ 7\%,
in good agreement with the discussion above.

\section{Non-uniformity of the detector response}
\label{nonuniform}

A further limitation to the effective S/N ratio is due to the 
non-uniformity of the sensor material properties and thickness, 
and to variations of the gain of the signal processing electronics.
Over sensor wafers with areas of the order of tens or hundreds 
of $\mathrm{cm}^2$ significant variations in process parameters 
are expected. Typically, these are relatively smooth variations. 
For instance, bulk doping variations are significant only at
scales of order one $\mathrm{mm}$, since the pulling 
of a crystal is a high temperature process and diffusion
smoothes out any abrupt non-uniformities. 
The spatial resolution, on the other hand, is only affected
by differences in response of neighbouring pixels. 
In this Section we estimate local and large-scale variations
in response of a DEPFET prototype. 

\begin{figure}[h!]
 \centering 
  \includegraphics[width=\columnwidth]{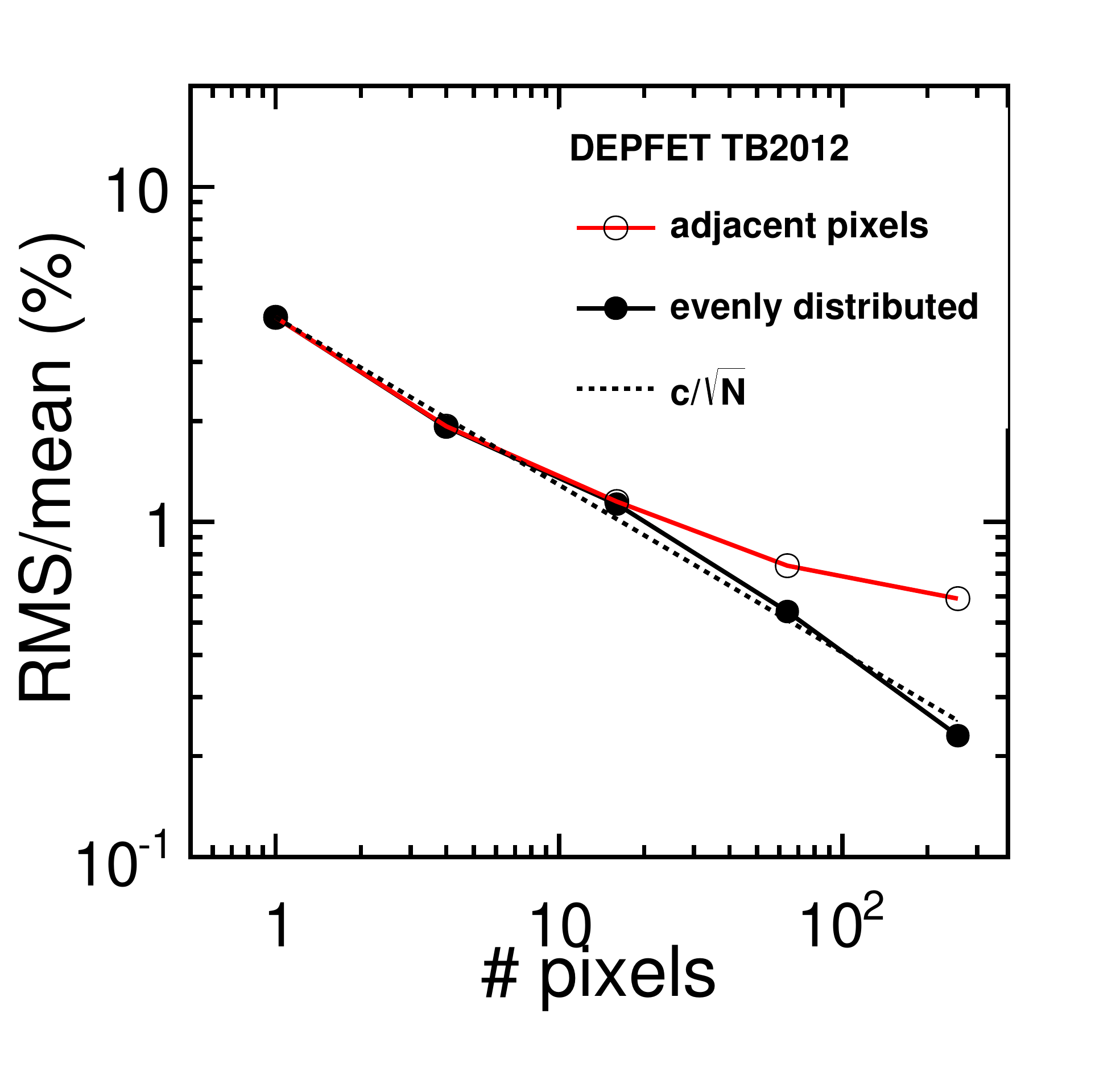}
 \caption{The Root-Mean-Square of the average response of groups of pixels versus the number of pixels included in the group. The red curves with open markers correspond to groups of adjacent pixels, while the black curve with filled markers corresponds to a groups that include pixels that are evenly distributed over the full sensor area.}
\label{fig:uniformity}
\end{figure}

Direct estimates in DEPFET beam tests of the size of pixel-to-pixel response variations 
have limited precision due to the available statistics.
The leftmost point in Figure~\ref{fig:uniformity} presents the 
Root-Mean-Square (RMS) of the response measured on each of the 2048 pixels 
of a small DEPFET prototype as the mean signal deposited by 
an average of approximately 50 MIP tracks. The observed RMS of approximately 4\% is
dominated by the statistical uncertainty on the response measurement
and must be interpreted as an upper limit on the true response variations.

Measuring the average response of groups of $N$ pixels (with $N=$ 4,16,64,256,
as indicated along the $x$-axis) the RMS variation decreases. 
Initially, we find the decrease follows the RMS $\propto 1/\sqrt{N}$ curve
that is indicated as dashed line. 
As soon as we reach the level of true response variations the observed
curve departs from the dashed line. For large groups of pixels the RMS is
found to depend on the exact scheme that is used to form groups of $N$ 
pixels. For groups of $\sqrt{N} \times \sqrt{N}$ adjacent pixels the average
remains sensitive to large-scale variations. The red curve with open markers
is indeed found to depart from $1/\sqrt{N}$ curve for RMS values slightly below
1\%. If the pixels of each group are evenly distributed over the sensor area 
wafer-scale variations cancel out in the RMS. The continuous black curve with 
filled markers continues to follow the dashed line down to an RMS of 0.2\%. 

This measurement shows that, even if detector responses are likely non-uniform 
at the \%-level over large areas and pixel counts, variations in the response of adjacent 
pixels can be controlled to a much higher degree. With a careful design and 
calibration scheme we are therefore optimistic that this limitation
can be circumvented.

\section{Summary and discussion}
\label{sec:conclusion}

In this paper we have explored the potential of silicon-based 
charged particle detectors that rely on interpolation of the
signal in adjacent sensor segments to achieve a spatial resolution
well below the size of the read-out cells.
We find that several limitations inherent in the physical process that is 
responsible for the signal limit the ultimate resolution that
can be obtained with this scheme. 

Using a combination of Monte Carlo simulation and test beam data we 
establish that energetic electrons forming secondary 
tracks known as $\delta$-electrons limit the resolution to the level
of approximately 1~\um{}, checking the improvement 
of the resolution with increasing S/N ratio predicted 
by~\cite{Turchetta:1993vu}. It seems impossible, therefore, to 
produce position-sensitive devices based
on interpolation of the signal in adjacent pixels or strips with a 
resolution significantly below 1~\um{}.

For particles that traverse the silicon detector under an angle fluctuations 
of the signal have an important impact on the spatial resolution that
becomes more pronounced as devices get thinner.
For the case considered here, where the projection 
of the particle trajectory on the read-out plane is equal to the read-out
pitch, the best spatial resolution is limited to 5\% of the pitch for
thick devices ($d=$450~\um) and to 9\% for thin devices ($d=$ 50~\um). 

Finally, we have evaluated pixel-to-pixel response variations. We find that,
unless special care is taken to ensure uniformity, large-area sensors
are likely to present variations in response of the order of 1\%.
The response of direct neighbours in the pixel matrix is, however, found
to be significantly more uniform. The impact on the spatial resolution
of such variations is therefore limited.

The results we have found apply to a broad range of detectors - 
including the state-of-the-art silicon $\mu$-strip
and pixel detectors employed in large-scale tracking systems in 
collider experiments. Clearly, solutions can be envisaged that
circumvent these limitations. In applications where multiple 
scattering in the detector material is of minor importance,
the stacked sensors of References~\cite{heijne, Field} can
achieve sub-micron precision. Ultra-thin sensors with very fine
pixel pitch (1 $\times$ 1 $\um^2$, with a thickness of 1~\um{}), when 
technologically feasible, would also overcome these limitations.

\bibliographystyle{IEEEtran}
\bibliography{limitations}{}
\vfill
\end{document}